\documentclass[sigconf, anonymous=false]{acmart}

\usepackage{booktabs} 
\usepackage{multirow}
\usepackage{sistyle}
\SIthousandsep{,}
\usepackage{bbm}
\usepackage[british]{babel}
\usepackage{pgfplots}
\usetikzlibrary{patterns}
\usepackage{filecontents}
\usepackage{array}
\newcolumntype{C}[1]{>{\centering\let\newline\\\arraybackslash\hspace{0pt}}m{#1}}
\newcolumntype{L}[1]{>{\raggedright\let\newline\\\arraybackslash\hspace{0pt}}m{#1}}
\newcolumntype{R}[1]{>{\raggedleft\let\newline\\\arraybackslash\hspace{0pt}}m{#1}}

\setcopyright{rightsretained}

\copyrightyear{2019} 
\acmYear{2019} 
\setcopyright{acmcopyright}
\acmConference[SIGIR '19]{Proceedings of the 42nd International ACM SIGIR Conference on Research and Development in Information Retrieval}{July 21--25, 2019}{Paris, France}
\acmBooktitle{Proceedings of the 42nd International ACM SIGIR Conference on Research and Development in Information Retrieval (SIGIR '19), July 21--25, 2019, Paris, France}
\acmPrice{15.00}
\acmDOI{10.1145/3331184.3331403}
\acmISBN{978-1-4503-6172-9/19/07}

\begin{document}
\fancyhead{}

\title{MatchZoo: A Learning, Practicing, and Developing System for Neural Text Matching}

\author{Jiafeng Guo$^{\dagger, \ddag}$, Yixing Fan$^{\dagger, \ddag}$, Xiang Ji$^{\ast}$ and Xueqi Cheng$^{\dagger, \ddag}$}
\affiliation{
  \institution{${\dagger}$University of Chinese Academy of Sciences, Beijing, China\\
  $^{\ddag}$CAS Key Lab of Network Data Science and Technology, Institute of Computing Technology,\\ Chinese Academy of Sciences, Beijing, China\\
  	$^{\ast}$Beijing Institute of Technology University, Beijing, China
  } 
}
\email{{guojiafeng, fanyixing, jixiang, cxq}@ict.ac.cn}


\begin{abstract}
Text matching is the core problem in many natural language processing (NLP) tasks, such as information retrieval, question answering, and conversation. Recently, deep leaning technology has been widely adopted for text matching, making neural text matching a new and active research domain. With a large number of neural matching models emerging rapidly, it becomes more and more difficult for researchers, especially those newcomers, to learn and understand these new models. Moreover, it is usually difficult to try these models due to the tedious data pre-processing, complicated parameter configuration, and massive optimization tricks, not to mention the unavailability of public codes sometimes. Finally, for researchers who want to develop new models, it is also not an easy task to implement a neural text matching model from scratch, and to compare with a bunch of existing models. In this paper, therefore, we present a novel system, namely MatchZoo, to facilitate the learning, practicing and designing of neural text matching models. The system consists of a powerful matching library and a user-friendly and interactive studio, which can help researchers: 1) to learn state-of-the-art neural text matching models systematically, 2) to train, test and apply these models with simple configurable steps; and 3) to develop their own models with rich APIs and assistance. 


\end{abstract}

%
%



\maketitle

{\fontsize{8pt}{8pt} \selectfont
\textbf{ACM Reference Format:}\\
Jiafeng Guo, Yixing Fan, Xiang Ji and Xueqi Cheng. 2019. Match-Zoo: A Learning, Practicing, and Developing System for Neural Text Matching. In 
\textit{Proceedings of the 42nd Int'l ACM SIGIR Conference on Research and Development in Information Retrieval (SIGIR'19), July 21--25, 2019, Paris, France.} 
ACM, NY, NY, USA, 4 pages. https://doi.org/10.1145/3331184.3331403 
}

\section{Introduction}
Many natural language processing (NLP) tasks can be formulated as a matching problem between two texts. For example, information retrieval is about the matching between a query and a document, question answering attempts to match an answer to a question, while conversation could be viewed as the matching between a response and an input utterance. In recent years, with the advance of deep learning technology, we have witnessed a growing body of work in applying shallow or deep neural models for the text matching problem, leading to a new and active research direction named neural text matching in this work.

Just as the emergence of Web applications leads to information overload, the quick growth of neural text matching models also brings some kind of ``model overload'' to researchers. Firstly, the learning cost increases significantly with the number of neural matching models. It becomes more and more difficult for researchers, especially those newcomers to this area, to learn and understand these new models. Secondly, it takes a lot of effort to try or apply existing models. Sometimes the public code of a specific model is not available. If it is available, it might be a stand-alone algorithm and you need to conduct tedious data pre-processing, complicated parameter configuration, and massive optimization tricks before you can apply it to your dataset. Finally, for researchers who want to develop new models, it is not an easy task either. It takes time to implement a neural text matching model from scratch, and even more time to compare with the a bunch of existing models.

In this demo, we present a novel system, namely MatchZoo, to tackle the above challenges. The system is designed to facilitate the learning, practicing and developing of neural text matching models. The overall architecture of the system consists of two major components: 1) the MatchZoo library: a neural text matching library which implements popular neural text matching algorithms as well as rich APIs for data pre-processing, model construction, training and testing, and automatic machine learning (AutoML); 2) the MatchZoo studio: a user friendly and interactive Web interface which enables users to browse, configure, run, test, apply and create neural text matching models.

With the MatchZoo system\footnote{http://www.bigdatalab.ac.cn/matchzoo/\#/}, researchers can: 1) learn state-of-the-art neural text matching models systematically, including the model descriptions, network structures, performances, as well as the code implementation; 2)  apply these models easily through simple parameter configuration, interactive training/testing, and direct application on real data; and 3) develop their own models rapidly with rich pre-processing APIs, off-the-shelf network layers, popular learning objectives/optimization methods/evaluation metrics, and fully-assistant notebooks. 

The MatchZoo system is built upon the previously released open source  toolkit \cite{fan2017matchzoo} with updated library and fresh new interfaces. There have been some related system in this direction, such as TFRank \cite{TensorflowRanking2018} and Anserini \cite{yang2017anserini}. However, TFRank only focuses on learning to rank techniques based on TensoFlow while Anserini is an IR toolkit on reproducing retrieval models. Our system is significantly different from them with a focus on helping researchers learning, practicing and developing neural text matching models.

\begin{figure}[tbp]
\centering
\includegraphics[scale=0.25]{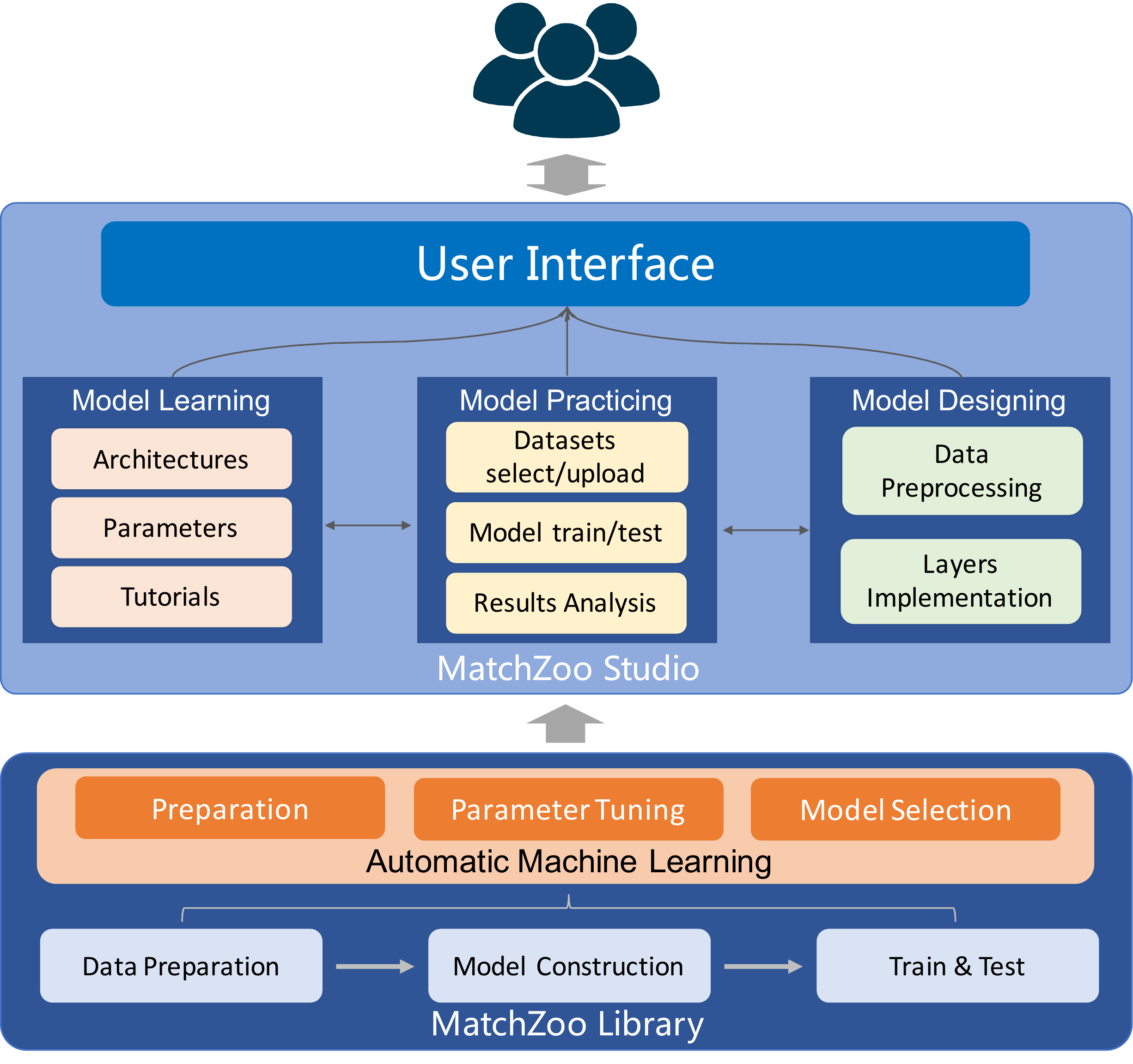}
\caption{An Overview of the System Architecture.}
\label{fig:architecture}
\end{figure}

\section{System Overview}
The architecture of the system is shown in the Figure \ref{fig:architecture}. The system consists of two major components, namely the MatchZoo library and the MatchZoo studio. 
The library provides a number of text processing units, popular neural text matching models, as well as matching based evaluation and loss functions, for all stages (i.e, data preparation, model construction, and train and test.) of the machine learning based text matching tasks. Moreover, we have also provided the AutoML operators to support automatic data preparation, hyper-parameter tuning, and model selection in the library.
The studio provides an interactive interface based on the MatchZoo library. There are three key functions, i.e., \textit{model learning}, \textit{model practicing}, and \textit{model designing}, to ease the process of learning, using and creating neural text matching models. The studio contains a user-friendly GUI which is built on the Web server, and users can interact with the studio through Web browsers.
 
 


\section{MatchZoo Library}
\begin{figure}[!t]
\centering
\includegraphics[scale=0.25]{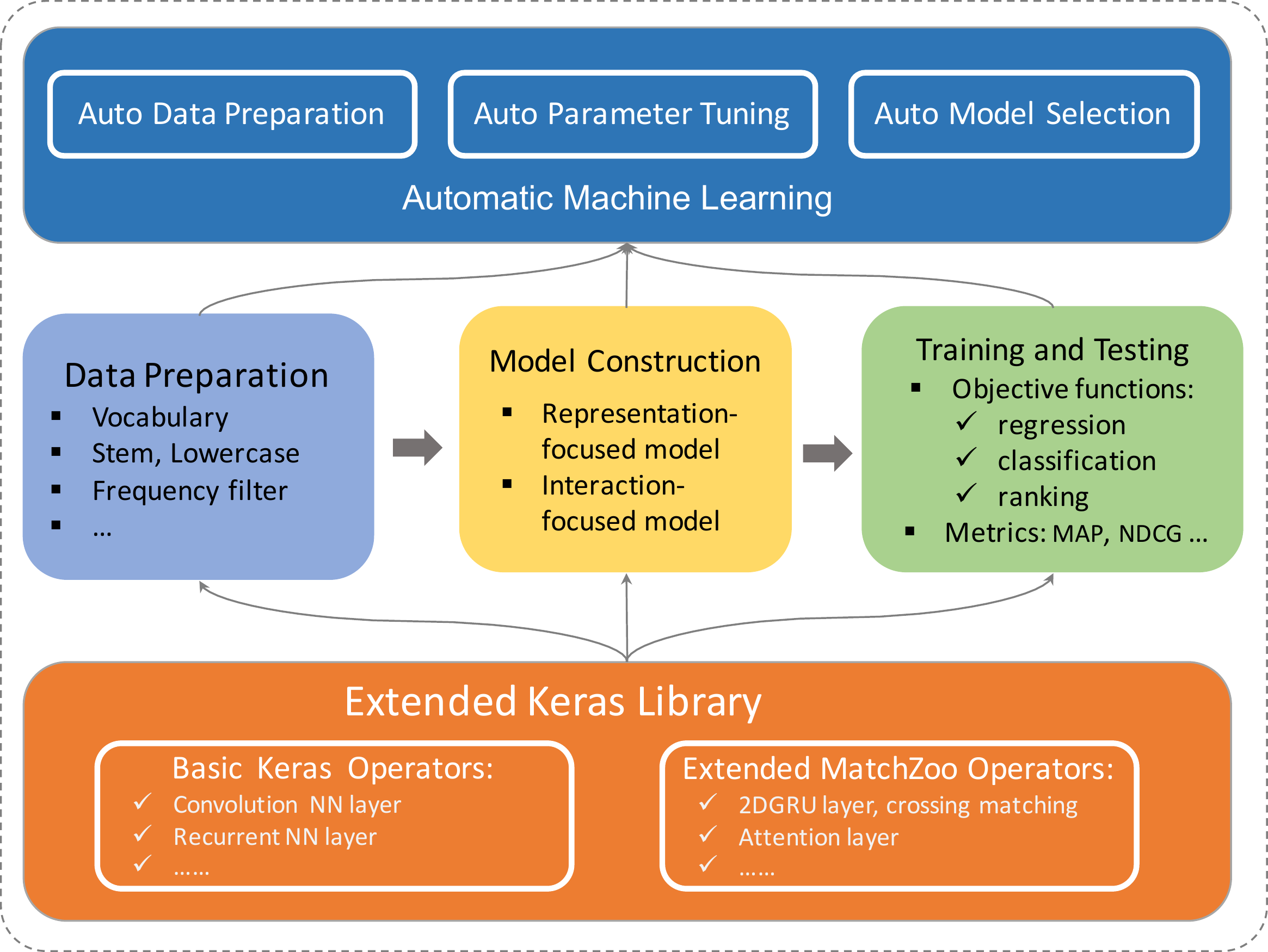}
\caption{An overview of the MatchZoo library.}
\label{fig:matchzoo}
\end{figure}

The MatchZoo library is to provide the functions supporting the high-level text matching tasks. Generally, the matching task can be decomposed into three steps, namely data preparation, model construction, and train/test. To support these steps, we extended the Keras library to include layer interfaces that are specially designed for text matching problems. Moreover, we have also added the automatic component in which the data preparation, hyper-parameter tuning, and model selection can be done automatically. This is very important as tuning machine learning hyper-parameters is a tedious yet crucial task, as the performance of an algorithm is highly dependent on the choice of hyper-parameters. In this way, we can largely alleviate the burden on tuning the hyper-parameters. 
The architecture of the MatchZoo library is shown in Figure \ref{fig:matchzoo}. 

\begin{figure*}[!t]
\centering
\includegraphics[scale=0.3]{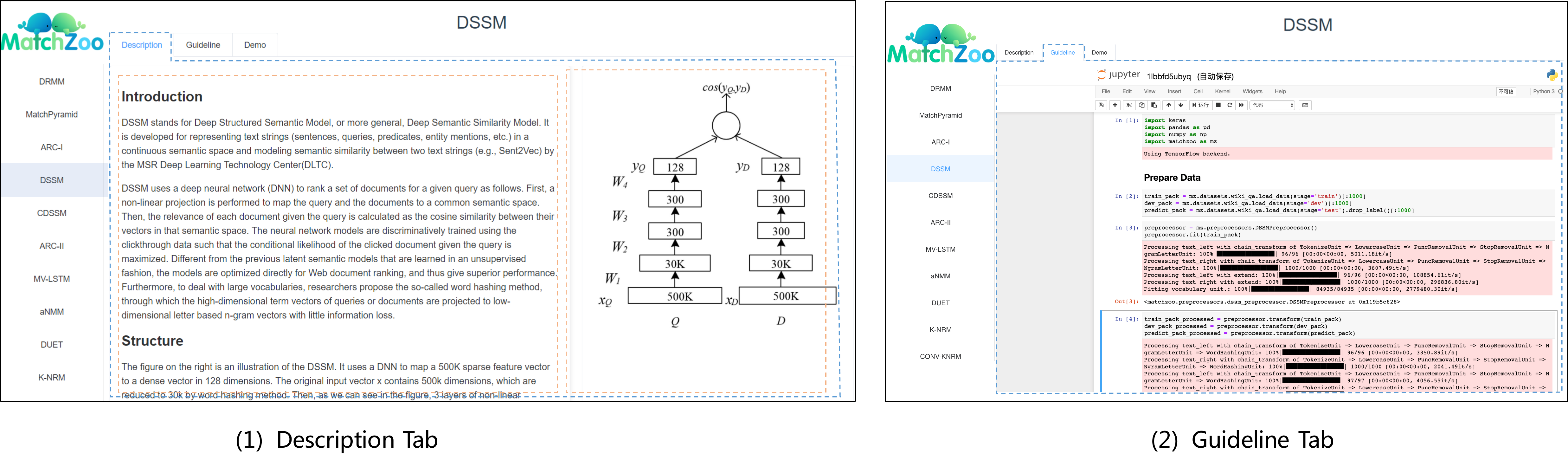}
\caption{The interface of the model learning component.}
\label{fig:model_learn}
\end{figure*}

\subsection{Data Preparation}
The data preparation module aims to convert the raw texts into the format of model's input. Here, we provided a number of text processing units where each unit is designed to perform a specific data transformation. 
Here, we list a few examples here.
\begin{itemize}
	\item \textbf{Lowercase Unit} converts all the characters into lower case.
	\item \textbf{FrequencyFilter Unit} filters out words based on pre-defined word frequency threshold.
	\item \textbf{PuncRemoval Unit} removes the punctuations from texts. 
	\item \textbf{Vocabulary Unit} transforms the word tokens into a sequence word indices.
	\item \textbf{WordHashing Unit} transforms the word tokens into tri-letter tokens.
\end{itemize}
All the processing units can be easily combined together to meet different model's data format requirement since they share a unified API. After converting the raw dataset to the desired format, the module provides three types of data batch modes, i.e., generating a batch of data in pointwise, pairwise or listwise manner.

\subsection{Model Construction}
In the model construction module, we employ Keras library to help users build the deep matching model layer by layer conveniently. The Keras library provides a set of common layers widely used in neural models, such as convolutional layer, pooling layer, dense layer and so on. To further facilitate the construction of deep text matching models, we extend the Keras library to provide some layer interfaces specifically designed for text matching. We list a few examples here.
\begin{itemize}
\item \textbf{Matching\_Matrix} layer builds a word-by-word matching matrix based on dot product, cosine similarity or indicator function~\cite{pang2016text}.
\item \textbf{Attention} layers builds a general attention layer for a pairs of text input.
\item \textbf{Matching\_Histogram} layer builds a matching histogram based on cosine similarity between word embeddings from two texts \cite{guo2016deep}.
\end{itemize}
Moreover, the library has implemented two schools of representative deep text matching models, namely representation-focused models and interaction-focused models~\cite{guo2016deep}. \begin{itemize}
\item \textbf{Representation-based models} include ARC-I~\cite{hu2014convolutional}, DSSM~\cite{huang2013learning}, CDSSM~\cite{shen2014learning}, MV-LSTM~\cite{wan2016deep}, and so on;
\item \textbf{Interaction-based models} include DRMM~\cite{guo2016deep}, ARC-II~\cite{hu2014convolutional}, KNRM \cite{xiong2017end}, and so on;
\end{itemize}
Users can apply these models out-of-the-box or modify them via simple configuration. 
\begin{figure}[!t]
\centering
\includegraphics[scale=0.3]{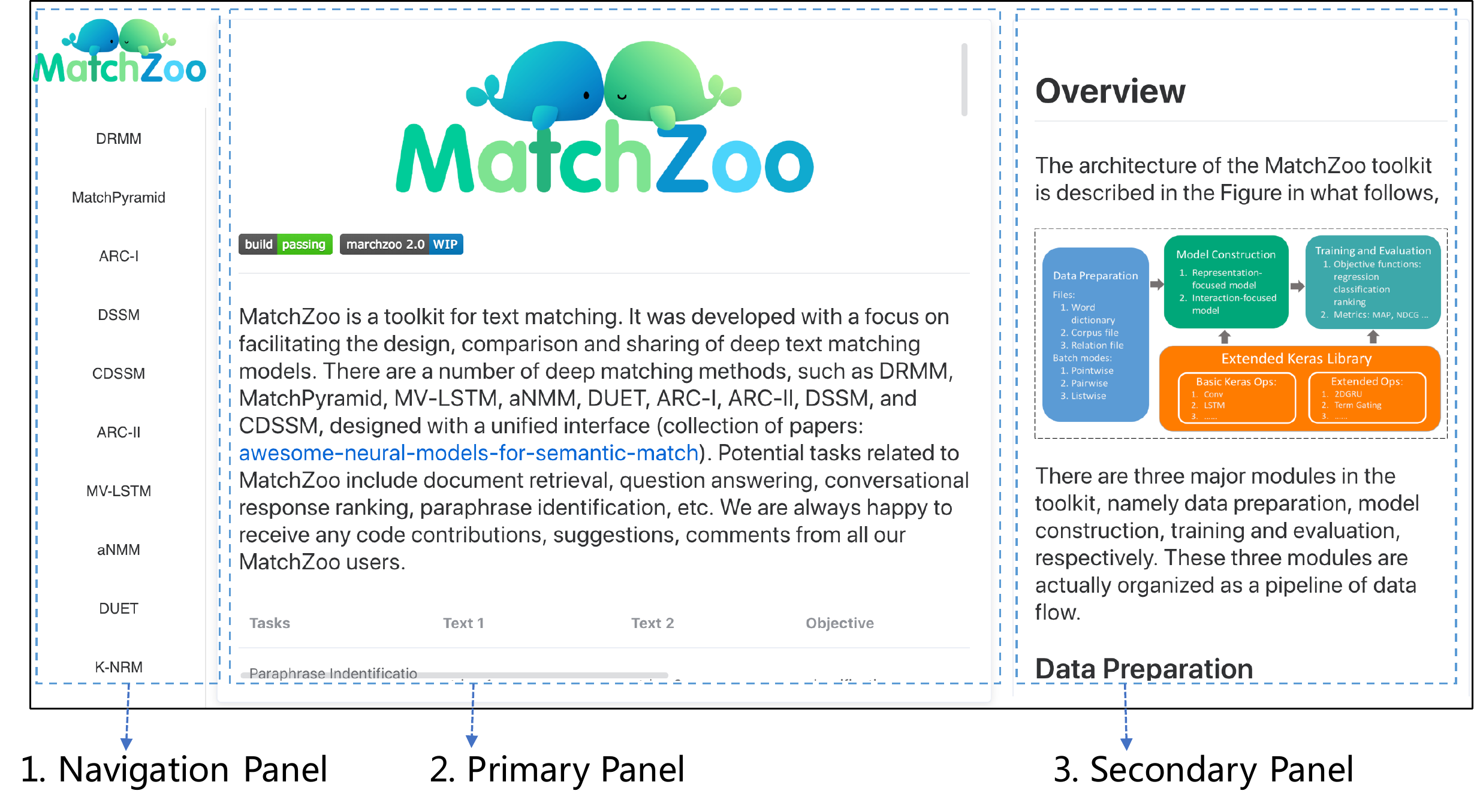}
\caption{The interface of the MatchZoo studio.}
\label{fig:studio}
\end{figure}

\subsection{Train \& Test}
For learning the deep matching models, the Library provides a variety of objective functions for regression, classification and ranking. For example, the ranking-related objective functions include several well-known pointwise, pairwise and listwise losses. It is flexible for users to pick up different objective functions in the training phase for optimization. 
For evaluation, the library provides several widely adopted evaluation metrics, such as Precision, MAP, and NDCG. 

\subsection{Automatic Machine Learning}
The AutoML component is to ease the application of neural text matching models by automatically conducting the data transformation, hyper-parameter tuning, and model selection. Specifically, each existing model is connected with a data transformer which directly converts the raw dataset into the required input format. To conduct AutoML, users just need to define the search space for all the hyper-parameters, then an automatic optimization process will be conducted through the random search algorithm \cite{bergstra2012random}. The best model will be selected according to some pre-determined metric. 

\section{MatchZoo Studio}
The MatchZoo studio provides a user-friendly Web GUI so that ordinary users can lean, practice, and develop neural text matching models easily. Figure \ref{fig:studio} shows an overview of the MatchZoo studio. As we can see, the interface is segmented into three vertical panels.
\begin{itemize}
    \item {Navigation panel} is on the left where users can select a neural matching model from the model list or choose to create a new model.   
    \item {Primary panel} is in the middle which includes three tabs namely description, guideline and train/test. These tabs are used to display the model description, interactive programming, and configurable experiments.
    \item {Secondary panel} is on the right which provides some auxiliary information such as detailed model structures, experimental results and API documentation.
\end{itemize}

\begin{figure*}[!t]
\centering
\includegraphics[scale=0.23]{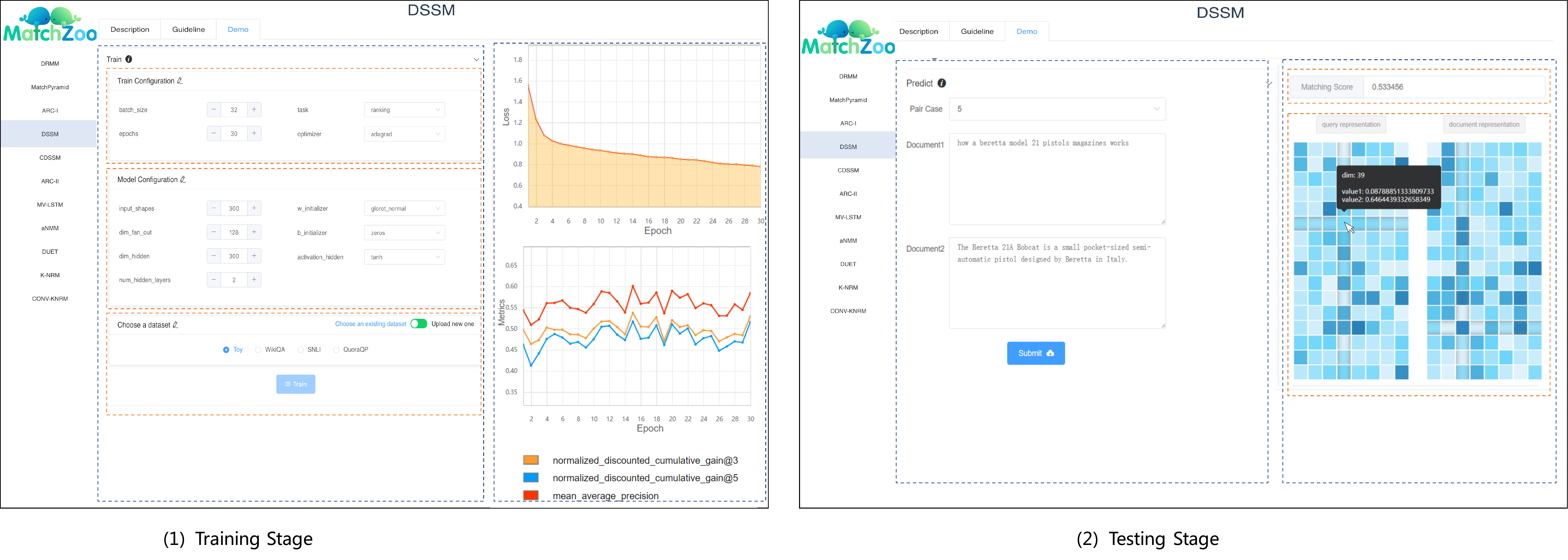}
\caption{The interface of the model practicing component.}
\label{fig:model_practice}
\end{figure*}

\subsection{Model Learning}
Figure \ref{fig:model_learn} shows the interface how users can learn different neural matching models in MatchZoo. Specifically, users can select a model in the navigation panel. Then, a systematical tutorial including theoretical descriptions and implementation details could be found under the description tab and guideline tab in the primary panel. As shown in Figure \ref{fig:model_learn}, the description tab contains a brief introduction of the model structure, parameters, performance of the selected neural text matching model DSSM. The guideline tab is an interactive Jupyter notebook. Under this tab, users can not only learn the original implementation code of DSSM, but also modify the code and experience with it.

\subsection{Model Practicing}
Figure \ref{fig:model_practice} shows the interface how users can practice different neural matching models in MatchZoo. After selecting a model from the navigation panel, there are two stages to experience with the model, namely training stage and testing stage. In training stage, as is shown in Figure \ref{fig:model_practice} (1), users can interactively configure the model hyper-parameters and select/upload a dataset in the primary panel. Then, the secondary panel will display the training process, including the loss curve on the training set and performance curves on the validation set. In testing stage, as is shown in Figure \ref{fig:model_practice} (2), users can type in or select two texts as inputs in the primary panel. Then, the secondary panel will show the matching score as well as the layer weights. Note here the example DSSM model is a representation-focused model, so the learned representation vector of the two inputs are displayed for comparison and intuitive understanding. For interaction-focused model, one can visualize the interaction matrix for model explanation.

\subsection{Model Designing}
\begin{figure}[!t]
\centering
\includegraphics[scale=0.28]{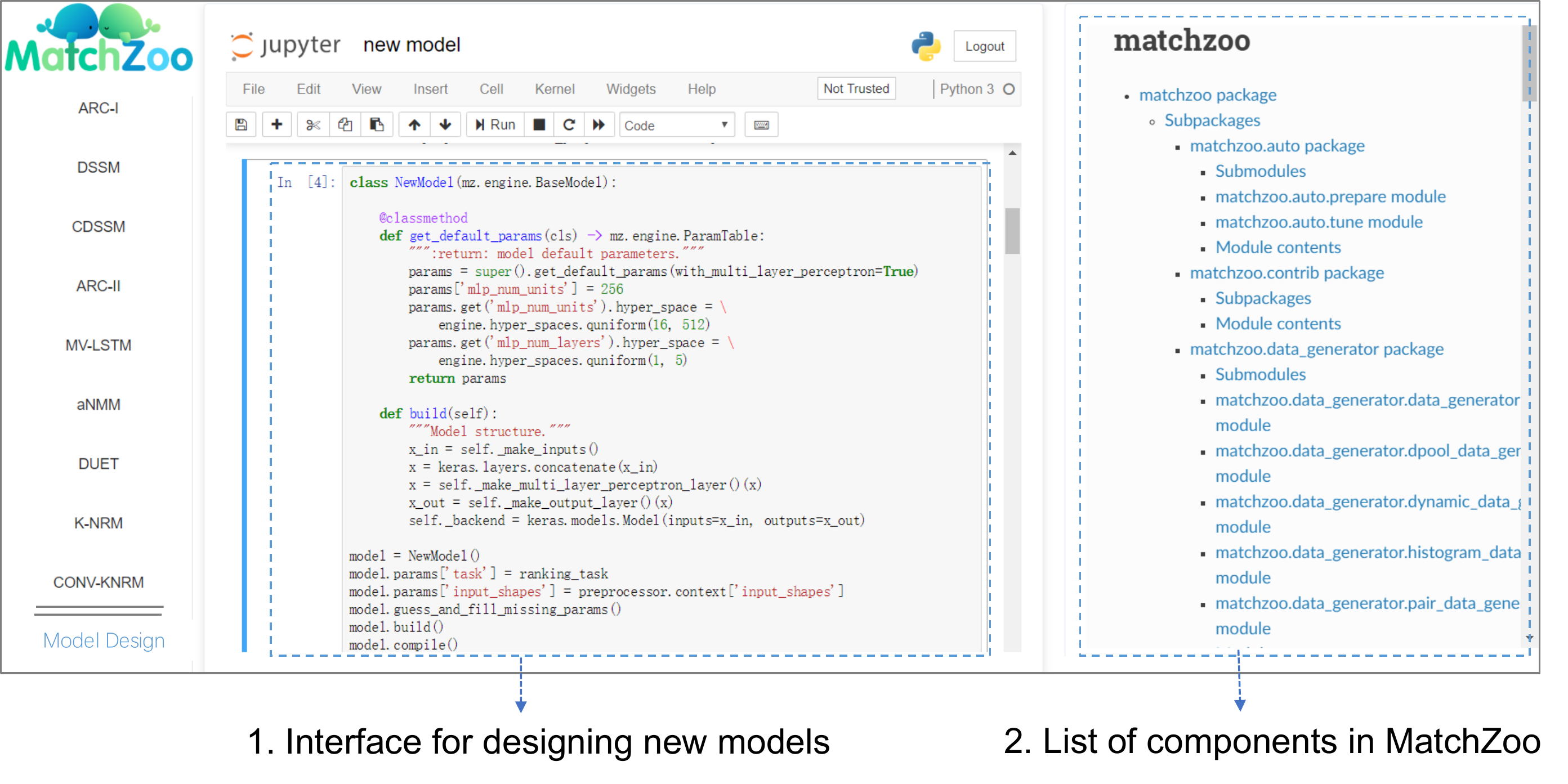}
\caption{The interface of the model designing component.}
\label{fig:model_design}
\end{figure}

Figure \ref{fig:model_design} shows the interface how users can create a new neural matching models in MatchZoo. Specifically, users can click the ``Model Design'' in the navigation panel. Then, a Jupyter Notebook will be present in the primary panel where users can directly implement his/her own neural matching model. At the same time, on the secondary panel, a detailed documentation about all the existing component APIs in MatchZoo would be displayed for users to search and access.

\section{Demo Plan}
We will present our system in the following aspects: (1) We will use a poster to give an overview of system architecture and briefly show the stages of the neural text matching process as well as the system components. (2) We will show the audience how to use the system to complete an example of text matching task, including data set pre-processing, model configuration, train, and test. (3) We will give a brief introduction of the neural text matching models in the system. (4) We will share our thoughts on the strengths and weakness of the system, and further discuss the future work.

\section{Acknowledgments}
This work was funded by the National Natural Science Foundation of China (NSFC) under Grants No. 61425016, 61722211, 61773362, and 61872338, the Youth Innovation Promotion Association CAS under Grants No. 20144310 and 2016102, the National Key R\&D Program of China under Grants No. 2016QY02D0405, and the Foundation and Frontier Research Key Program of Chongqing Science and Technology Commission (No. cstc2017jcyjBX0059).

\bibliographystyle{ACM-Reference-Format}


\end{document}